# Kinesin Motors and the Evolution of Intelligence

J. C. Phillips

Dept. of Physics and Astronomy, Rutgers University, Piscataway, N. J., 08854

Abstract

Intelligence is often discussed in terms of neural networks in the cerebral cortex, whose evolution has presumably been influenced by Darwinian selection. Here we present molecular evidence that one of the many kinesin motors, Kif14, has evolved to exhibit special features in its amino acid sequence that could have evolved to improve neural networks. The improvement is quantified by comparison of Kif14 sequences for 12 species. The special feature is level sets of hydrophobic extrema in water wave profiles based on several hydropathic scales. The most effective scale is a new one based on fractals, indicative of approach of globular curvatures to self-organized criticality.

There are at least 14 kinesin motor families [1,2], and the mechanics of the hand-over-hand kinesin step dragging cargo along tubulin have been well studied [3-5]. A bioinformatic survey of 1624 putative kinesins identified three families that are very widespread among species: Kif 1, 5, and 14 [6] (see their Fig. 1). According to Uniprot, Kif1 provides "anterograde axonal transport of synaptic vesicle precursors", Kif5 is "required for slow axonal transport of neurofilament proteins", and "during late neurogenesis, KIF 14 regulates the cerebellar, cerebral cortex and olfactory bulb development through regulation of apoptosis, cell proliferation and cell division (by similarity)". Other kinesins are less common and appear to be less involved in the complex task of neural network assembly and maintenance. All of the motor families except Kif14 walk towards the plus end [2].

Many distinctive features connecting protein sequences to their functions and evolution are related to globular-shaping water waves obtained by thermodynamic scaling. The general method used here is only twelve years old, is interdisciplinary, and is little known. It is reviewed in our earlier companion article on actin [7]. Actin polymers are also part of the cytoskeleton and are only 16% smaller than tubulin. A simple way of thinking about these scaling methods is that they are an upscale version of BLAST, which is now ~ 40 years old. During that time the protein data base has grown enormously and we are now in the genomic age.

Second-order phase transitions are characterized in principle by fractal (non-integral) exponents, which are difficult to measure. The discovery of 20 universal amino-acid specific fractals in the solvent accessible surface areas (SASA) of proteins [8,9] is by far the most important application of phase transition theory in the 21st century and the latter part of the 20th century. It enables extraction of large amounts of information from the vast array of protein sequences now available using high-precision methods that achieve thermodynamic accuracy. The fractals themselves are fixed by the survey of SASA for more than 5000 protein segments and are not adjustable parameters.

Phylogenetics counts numbers of identical or similar amino acids at specific sites using BLAST, and it is limited by the restriction to single sites. There is an alternative to the single site methods, which has Darwinian selectivity as an implicit feature, as corroborated by the identification of universal self-organized criticality in the solvent-accessible surface areas (SASA) of > 5000 protein amino acid segments from the modern Protein Data Base [8,9]. The lengths of the small segments $L = 2N + 1$ varied over a wide range from 3 to 45, but the interesting range turned out to be $M< = 9 \le L \le 35 = M>$. Across this common range they found linear behavior on a log-log plot (a power law, hence self-similar or scale-free) for each of the 20 amino acids centered on a given segment

$$\log SASA(L) \sim \text{const} - \Psi(aa) \log L \qquad (9 \le L \le 35)$$

Here $\Psi(aa)$ is a hydropathicity parameter. It arises because the longer segments fold back on themselves, occluding the SASA of the central aa. The most surprising aspect of this self-similar folded occlusion is that it is nearly universal on average across the proteome, and almost independent of the individual protein fold. This is a dramatic demonstration of the power of Darwinian selectivity involved in aqueous shaping of globular proteins, as discussed in detail elsewhere [10]. Moreover, the segmental character of the new scale [8,9] has a Darwinian echo: for each protein family one can identify an optimized sliding window width W*, over which $\Psi(aa)$ is best averaged to maximize evolutionary improvements; this averaged profile is denoted by Ψ(aa,W*). Profiles of Ψ(aa,W*) display the functional features that are being optimized by evolution, often involving modular (segmental) exchange [10].

Evolution has leveled sets of (more often) hydrophobic and (less often) hydrophilic extrema in profiles of many proteins [11-13]. Such leveling synchronizes protein multi-domain hydrodynamics, in accordance with Sethian's level set theory [14-16]. Synchronization is a common phenomenon and occurs in many scale-free networks, including social networks [17]. Protein-protein interactions are complex, and level sets are not the rule. However, when they do occur, they can be used to identify more easily quantified protein dynamics, and monitor functional evolution. In contrast to molecular dynamics simulations, it is easy to analyze profiles even of very large proteins of more than 1000 amino acids [11] near equilibrium. Here we inspect Kif 1,5 and 14 profiles and find that there are no level sets in the extrema of Kif1 profiles, while the Kif14 human profile exhibits striking level sets of hydrophobic extrema. Kif5 is an intermediate case, with only some level set aspects, so we focus on Kif 14 and its evolution in later species.

The optimal value of the sliding window width W* can be determined in different ways, depending on the protein family under study. Usually one can maximize differences between sequences of different species by finding extrema of variances of Ψ(aa,W) at W = W* (examples are [7, 11-13}. The Kif14 variance ratios decrease monotonically with W, so this technique fails. Here we studied graphs of human profiles with W ranging from 7 to 21, all of which showed level sets which became more level with evolution. Now maximizing the difference in leveling at the human value gave W* = 9. In general smaller values of W give higher resolution, while larger values of W are more likely to treat segmental dynamics more accurately. It is striking that this vale of W* = 9 coincides with the lower limit M< of the MZ fractal range. In addition to the MZ 2007 fractal scale, which is consistent with second-order phase transition and self-organized criticality [9,14], we also used the standard 1982 first-order phase transition (protein unfolding) hydropathicity scale (KD), which gave weaker results [18].

The effects of evolution on Kif14 are illustrated in Fig. 1, which compares human and chicken profiles. The chicken level set has 3 peaks which grow to 5 peaks in human. Alternatively, one can find the deviation from the mean for the five highest hydrophobic peaks. This is smallest in humans, and it increases for earlier species. The mean deviation of the highest five peaks for human Kif1 is 10.8, or 7 times larger than human Kif14 (practically not a level set). (Kif1 is the most common kinesin, and supports growth of cells in many tissues.) Altogether this is good circumstantial evidence that Kif14 is important to building and refining neural networks.

Numerical results for mean deviations of the highest five peaks of many species are shown in Table I, together with the number of BLAST positives that differ from human. The reader may judge which column more accurately reflects relative intelligence. Domestic dog breeds are characterized by an unrivaled diversity of morphologic traits and breed-associated behaviors resulting from human selective pressures. A recent survey of whole genome sequences of 722 dogs found 16 phenotypes that explain greater than 90% of body size variation in dogs [19,20]. If Kif14 motor sequences could be extracted from this data base, a correlation with canine intelligence variations could be discovered.

Readers who are unfamiliar with scale-free networks may be interested in a survey of a few of the many examples of synchronized level sets that occur in social and many other self-organized networks [17]. To the author's knowledge, proteins are the only naturally occurring system that exhibit synchronized level sets. These were discussed in abstract hydrodynamic or elastic contexts [14], amusingly similar to water at protein interfaces. The similarity to social networks emphasizes the importance of synchronization and level sets for understanding the origins of the last stages of protein evolution. There is no precedent for 20 fractals, and the key role played by fractals in science is far-reaching [17, 21].

## References

1. Lawrence, C. J.; Dawe, R. K.; Christie, K. R.; et al. A standardized kinesin nomenclature. J. Cell Biol. **167**, 19-22 (2004).
2. Miki, H.; Okada, Y.; Hirokawa, N. Analysis of the kinesin superfamily: insights into structure and function. Trends Cell Biol. **15**, 467-476 (2005).
3. Yildiz, A; Tomishige, M; Vale, RD; et al. Kinesin walks hand-over-hand. Science **303**, 676-678 (2004).
4. Carter, N. J.; Cross, R. A. Mechanics of the kinesin step. Nature **435**, 308-312 (2005).

5. Block, S. M. Kinesin motor mechanics: Binding, stepping, tracking, gating, and limping. Biophys. J. 92, 2986-2995 (2007).
6. Wickstead B1, Gull K, Richards TA. Patterns of kinesin evolution reveal a complex ancestral eukaryote with a multifunctional cytoskeleton. BMC Evol Biol. **10**, 110 (2010).
7. Moret, M. A.; Zebende, G. F.; Phillips, J. C. Beyond phylogenetics: Darwinian evolution of actin. Rev. Mex. Ingen. Biomed. **40**, 1-11 (2019).
8. Moret, M. A.; Zebende, G. F. Amino acid hydrophobicity and accessible surface area. Phys. Rev. E **75**, 011920 (2007).
9. Phillips, J. C Scaling and self-organized criticality in proteins: Lysozyme *c*. Phys. Rev. E **80**, 051916 (2009).
10. Phillips, J. C. Quantitative molecular scaling theory of protein amino acid sequences, structure, and functionality. arXiv 1610.04116 (2016).
11. Allan, D. C., Phillips, J. C. Evolution of the ubiquitin-activating enzyme Uba1 (E1). Phys. A **483**, 456-461 (2017).
12. Phillips, J. C. Self-organized networks with long-range interactions: tandem Darwinian evolution of α and β tubulin Proc. Nat. Acad. (USA) **117**, 7799-7802 (2020).
13. Phillips, J. C. Self-organized networks: Darwinian evolution of dynein rings, stalks and stalk heads. Proc. Nat. Acad. (USA) **117**, 7799-7802 (2020).
14. Saye, R. I.; Sethian, J. A. The Voronoi implicit interface method for computing multiphase physics. Proc. Nat. Acad. Sci. (USA) **49**, 19498-19503 (2011).
15. Chiappori, P.-A.; McCann, R. J.; Pass, B. Multi-to one-dimensional optimal transport. Comm. Pure Appl. Math. **70**, 2405-2444 (2017).
16. Niu, X.; Vardavas, R.; Caflisch, R. E.; et al. Level set simulation of directed self-assembly during epitaxial growth. Phys. Rev. B **74**, 193403 (2006).
17. Albert, R.; Barabasi, A.-L. Statistical mechanics of complex networks. Rev. Mod. Phys. **74,** 47-97 (2002).
18. Kyte, J.; Doolittle, R. F. A simple method for displaying the hydropathic character of a protein**.** J. Mol. Biol. **157,** 105-132 (1982).

19. Plaissais, J.; Kim, J.; Davis, B. W.; et al. Whole genome sequencing of canids reveals genomic regions under selection and variants influencing morphology. Nature Comm. **10**, 1489 (2019).
20. Li, Y.; vonHoldt, B. M.; Reynolds, A.; et al. Artificial Selection on Brain-Expressed Genes during the Domestication of Dog. Mol. Biol. Evol. **30**, 1867-1876 (2013).
21. Stanley, E.; Taylor, E. Fractals in Science. Springer (1994).

| Species | Mean Devia. | Positives |
|---|---|---|
| Human | 1.50 | 0 |
| Whale | 1.78 | 30 |
| Zfish | 1.88 | 40 |
| W Turkey | 2.52 | 48 |
| Elephant | 2.55 | 10 |
| Mouse | 2.73 | 17 |
| Pol. Bear | 2.76 | 10 |
| Rabbit | 2.94 | 15 |
| Fox | 3.06 | 14 |
| Horse | 3.42 | 21 |
| Chicken | 3.91 | 45 |
| Mole Rat | 4.10 | 12 |

\\

Table I. Deviations from the mean for each species by the five highest hydrophobic peaks in the Kif14 motor profiles. Also shown are the number of BLAST positives that differ from human. The small values of mean deviation for whale, zebra fish, and wild turkey may represent the effects of extreme environmental conditions. Such anomalies could be minimized in canine data.

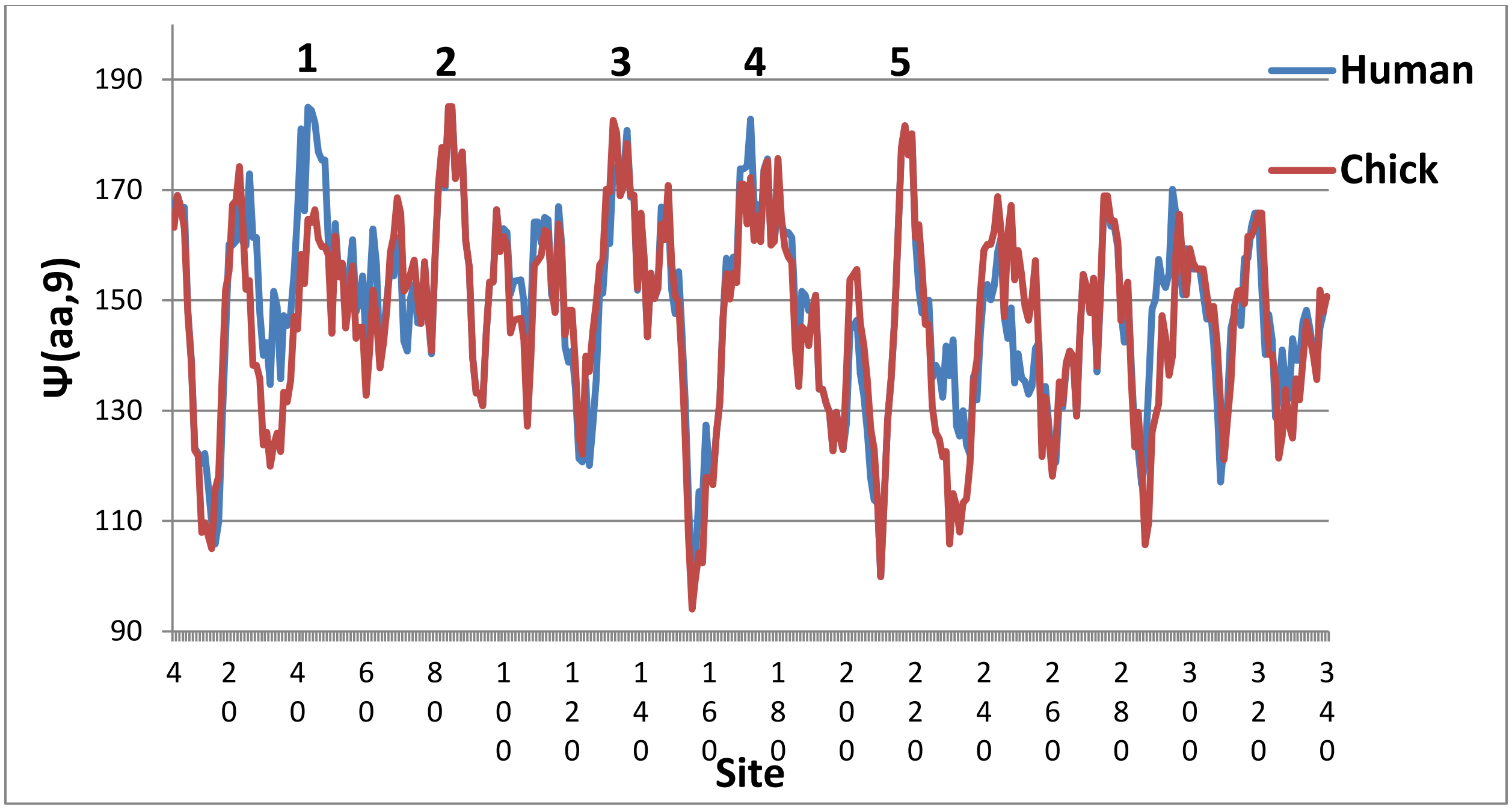

Fig. 1. Evolution leveled hydrophobic peaks 2, 3 and 5 in chicken, and further evolution added level peaks 1 and 4 in human. The reader may also observe partial leveling in chickens and humans of the five lowest hydrophilic minima near 10, 150, 210, 230 and 290.